\documentclass[onecolumn, 12pt, draftclsnofoot, comsoc]{IEEEtran}

\usepackage{amsmath,amssymb,amsfonts}
\usepackage{algorithm} 
\usepackage{algcompatible}
\usepackage{algpseudocode}

\usepackage[bookmarks,colorlinks]{hyperref} 
\hypersetup{colorlinks,citecolor= red,filecolor= blue,linkcolor= blue,urlcolor=blue}

\usepackage{graphicx,subfigure}%
\usepackage[usenames,dvipsnames]{xcolor}
\usepackage{verbatim}
\usepackage{soul}
\definecolor{LightBlue}{rgb}{0.75,0.936,1.00}
\sethlcolor{LightBlue}

\usepackage{tikz,pgfplots}

\usepackage[printonlyused,withpage]{acronym}
\usepackage{cite} 
\usepackage{color}
\usepackage{comment} 

\usepackage{mathrsfs} 
\usepackage{scalerel} 
\usepackage[T3,T1]{fontenc} 
\DeclareSymbolFont{tipa}{T3}{cmr}{m}{n}
\DeclareMathAccent{\invbreve}{\mathalpha}{tipa}{16}
\usepackage{psfrag}
\usepackage{multirow}
\usepackage{makecell}

\begin{document}

\title{Rate-Splitting Multiple Access for GEO-LEO Coexisting Satellite Systems: A Traffic-Aware Throughput Maximization Precoder Design}

\author{Jaehak Ryu, Aryan~Kaushik, Byungju~Lee, and Wonjae~Shin
    \thanks{
        J.~Ryu and W.~Shin are with the School of Electrical Engineering, Korea University, Seoul, South Korea.
        A.~Kaushik is with the School of Engineering and Informatics,  University of Sussex, Brighton, United Kingdom.
        B.~Lee is with the Department of Information and Telecommunication Engineering, Incheon National University, Incheon, South Korea.
        (\textit{Corresponding authors: B.~Lee and W.~Shin}).
	}
}

\maketitle




\begin{abstract}
The frequency coexistence between geostationary orbit (GEO) and low earth orbit (LEO) satellite systems is expected to be a promising approach for relieving spectrum scarcity.
However, it is essential to manage mutual interference between GEO and LEO satellite systems for frequency coexistence.
Specifically, \emph{in-line interference}, caused by LEO satellites moving near the line-of-sight path between GEO satellite and GEO users (GUs), can significantly degrade GEO system throughput.
This paper put forth a novel rate-splitting multiple access (RSMA) with a super-common message for GEO-LEO coexisting satellite systems (CSS).
By employing a super-common message that GUs can decode, GUs can mitigate the in-line interference by successive interference cancellation (SIC).
%
Moreover, we formulate a traffic-aware throughput maximization (TTM) problem to satisfy the heterogeneous traffic demands of users by minimizing total unmet throughput demands (or user dissatisfaction).
By doing so, the TTM precoder can be flexibly adjusted according to the interference leakage from LEO satellites to GUs and target traffic demands. 
Numerical results confirm that our proposed method ensures seamless connectivity even in the GEO-LEO in-line interference regime under imperfect channel state information (CSI) at both the transmitter and receiver.
\end{abstract}



\section{Introduction}\label{sec:intro}
{With the development of sixth-generation (6G) wireless networks, satellite communication (SATCOM) has sparked significant interest in providing massive connectivity \cite{Howon:JourCommunNet23}.
In particular, low earth orbit (LEO) satellites, which offer high throughput and small propagation delay compared to geostationary orbit (GEO) satellites, are rapidly emerging as key providers of global coverage.
However, the ongoing expansion of LEO satellites with the advancement of SATCOM faces a significant challenge due to limited spectrum resources. 
The frequency coexistence of the LEO system with the existing GEO system is expected to play a crucial role in alleviating the spectrum scarcity \cite{Christophe:DySPAN19,Wang:ICCC18,Ting:Int-J-SatCom-Net21,Yunfeng:Access20}.
It is important to note that this frequency coexistence approach introduces mutual interference, which could degrade the system throughput for GEO-LEO coexisting satellite systems (CSS).}
Specifically, as LEO satellites move near a line-of-sight path between the GEO satellite and the GEO user (GU), the GU can receive severe interference from LEO satellites.
This is called \emph{in-line interference}, which can disrupt the stable operation of the existing GEO system.
Note that the International Telecommunications Union (ITU) regulates the LEO system to protect the existing GEO system from unacceptable interference for GEO-LEO CSS \cite{ITU_R:Article}. 

Previous studies have proposed various types of interference avoidance approaches to prevent in-line interference from LEO satellites to GUs \cite{Wang:ICCC18,Ting:Int-J-SatCom-Net21,Yunfeng:Access20}.
One interference avoidance strategy is implementing an exclusive zone, which prohibits LEO satellites from providing services in the in-line interference regime \cite{Wang:ICCC18}.
Also, the progressive pitch scheme for OneWeb was discussed in \cite{Ting:Int-J-SatCom-Net21}, involving adjustments to LEO satellite attitude and reducing the number of spot beams to attenuate the interference leakage (IL) from LEO satellites to GUs in the in-line interference regime.
The above strategies can protect the GEO system at the cost of frequent coverage holes for LEO constellation systems.
Moreover, cognitive radio technology prioritizes frequency band allocation to the GEO system, followed by the LEO system, to prevent unacceptable IL from LEO satellites to GUs \cite{Yunfeng:Access20}. 
On the other hand, this approach constrains frequency resources available to the LEO system in the in-line interference regime, degrading its service quality.
Despite most studies focusing on protecting the GEO system, this, in turn, leads to reduced coverage for the LEO system.

Recently, rate-splitting multiple access (RSMA) has emerged as the flexible interference management framework that provides high spectrum efficiency and robustness against imperfect channel state information (CSI) \cite{Yijie:Tutor22,Khan:TWC23,Byungju:TVT23,Jeonghun:NetworkMag23,Huanxi:WCL23}.
The authors in \cite{Khan:TWC23} designed the RSMA power allocation to maximize the sum rate for all users with perfect CSI. 
In \cite{Byungju:TVT23}, RSMA with the max-min fairness (MMF) metric was proposed for both imperfect CSI at the transmitter (CSIT) and CSI at the receiver (CSIR) to maximize the minimum rate among all users.
However, considering the broad coverage of satellite, traffic demands in spatial distribution and temporal variation may be uneven. 
Thus, there is a need to develop a new metric designed to fulfill heterogeneous traffic demands among users while maximizing overall system throughput and adapting to channel conditions within the finite power budget \cite{Huanxi:WCL23,Abdu:OJCOM22}.

This paper proposes traffic-aware throughput maximization (TTM) based on RSMA with the super-common message for GEO-LEO CSS.
The main contributions of this paper are summarized as follows. 

\begin{itemize}
\item We put forth a novel RSMA framework with a super-common message for GEO-LEO CSS to manage the in-line interference.
Compared to existing avoidance strategies that restrict the LEO system \cite{Wang:ICCC18,Ting:Int-J-SatCom-Net21,Yunfeng:Access20}, our framework enables in-line interference to be flexibly decoded.
This is achieved through a super-common message that can be initially decoded and removed by all GUs and LEO users (LUs) via successive interference cancellation (SIC) (see Fig. 1).
By allocating power to a super-common message according to the dynamic level of in-line interference, we can ensure seamless throughput for GEO-LEO CSS.
\item To address the challenge of heterogeneous traffic demands among users, we formulate the TTM problem. This problem aims to satisfy each user's traffic demand by minimizing unmet throughput (or user dissatisfaction), which denotes the unfulfilled portion of the users' traffic demand within the given power budget for the satellite.
\item Given the non-convex nature of the original TTM problem, we employ an alternating optimization algorithm based on semidefinite relaxation (SDR) and concave-convex procedure (CCCP) to convert the problem into a convex form.
Subsequently, although we derive candidate solutions that satisfy the rank-1 constraint, they may not meet the IL constraint for GUs and the total power constraint.
To address these issues, we propose an algorithm that includes rescaling the candidate solutions to ensure a feasible solution.

\item Through numerical results, the proposed method shows seamless connectivity even in the GEO-LEO in-line interference regime under both imperfect CSIT and CSIR.
\end{itemize}

\emph{Notations}: $\mathbb{E}(\cdot)$ denotes the expectation operation. ${\sf tr}(\cdot)$ is the trace operation. $\odot$ is the Hadamard product. $J_1(\cdot)$ and $J_3(\cdot)$ are the first and third-order Bessel functions of the first kind, respectively. $\mathbf{I}_N$ indicates a $N$ by $N$ identity matrix. \textrm{blkdiag}($\cdot$) indicates a block-diagonal operation.


\section{System Model}\label{SEC:system-model}
We consider GEO-LEO CSS, in which each GU and LU communicates with its target satellite over the same spectrum of resources.
The GEO satellite with $N_{\sf{G}}$ antenna feeds and the LEO satellite with $N_{\sf{L}}$ antenna feeds serve $K_{\sf{G}}$ GUs and $K_{\sf{L}}$ LUs, respectively.
$\mathcal{K_G} \triangleq \{1,2,\dotsc,K_{\sf{G}}\}$ and $\mathcal{K_L} \triangleq \{1,2,\dotsc,K_{\sf{L}}\}$ are index sets for GUs and LUs, respectively. 
Each user has a single antenna, and each satellite employs a single feed per beam (SFPB) architecture where the number of antenna feeds is identical to the number of adjacent beams. 
We assume that the GEO satellite employs a frame-based transmission structure, i.e., multicast transmission, following DVB-S2X to serve all GUs within each beam \cite{Huanxi:WCL23}, while the LEO satellite wants to convey independent data streams to each LU, i.e., unicast transmission.
Moreover, we consider the user-overloaded scenario where the number of LUs exceeds the number of LEO satellite antenna feeds ($K_{\sf{L}} > N_{\sf{L}}$).

\subsection{Satellite Channel Model} 
The channel vector between the LEO satellite and $k_{l}$-th LU can be expressed as \cite{Ahmad:Doctoral16},
{
\begin{equation}
\mathbf{h}_{k_{l}}=\sqrt{\left(\frac{\lambda}{4 \pi d_{{\sf{L}},k_{l}}}\right)^2 \frac{G_{\sf{R}}^{\sf{max}}}{\kappa B T}} \mathbf{b}_{{\sf{L}},k_{l}}^{\frac{1}{2}} \odot e^{j {\phi}_{{\sf{L}},k_{l}}},
\end{equation}}where $\left(\lambda / 4 \pi d_{{\sf{L}},k_{l}}\right)^2$ is the effect of free space path loss between LEO satellite and $k_{l}$-th LU, $\kappa$ is the Boltzmann's constant, $B$ is the noise bandwidth and $T$ is the receiver noise temperature. $G_{\sf{R}}^{\sf{max}}$ is the maximum receive antenna gain, assuming that each user continuously tracks the target satellite to maintain the maximum receive antenna gain. $\mathbf{b}_{{\sf{L}},k_{l}} = [b_{k_l,1},\ldots,b_{k_l,N_{\sf{L}}}]^T \in \mathbb{C}^{N_{\sf{L}} \times 1}$ is radiation pattern for $N_{\sf{L}}$ antenna feeds. The transmit antenna gain between the $n_{l}$-th feed of the LEO satellite and $k_{l}$-th LU is expressed as
{
\begin{equation}
b_{k_l,n_l}=G_{\sf{T}}^{\sf{max}} \left[\frac{J_1(u_{k_l,n_l})}{2u_{k_l,n_l}} + \frac{36 J_3(u_{k_l,n_l})}{{u_{k_l,n_l}}^3}\right]^2,
\end{equation}}where $G_{\sf{T}}^{\sf{max}}$ is the maximum transmit antenna gain and $u_{k_l,n_l}=2.07123\left({\textrm{sin}}(\varphi_{k_l,n_l})/{\textrm{sin}}(\varphi_{{\sf{L}},{\textrm{3dB}}})\right)$. 
We denote the angle between the $k_l$-th LU and the $n_l$-th LEO beam as $\varphi_{k_l,n_l}$, and the ${\textrm{3dB}}$ angle for the $n_l$-th LEO beam as $\varphi_{{\sf{L}},\textrm{3dB}}$, respectively. ${\phi}_{{\sf{L}},k_{l}} \triangleq \left[\phi_{k_l,1},\dotsc, \phi_{k_l,N_{\sf{L}}} \right]^T \in \mathbb{C}^{N_{\sf{L}} \times 1}$ denotes the channel phase components for the $k_l$-th LU. 
Considering the imperfect CSI, the actual channel between LEO satellite and $k_l$-th LU can be formulated as $\mathbf{h}_{k_l} = \hat{\mathbf{h}}_{k_{l}} + \mathbf{e}_{{\sf{L}},k_l}$, where $\hat{\mathbf{h}}_{k_{l}}$ is the estimated channel and the channel estimation error $\mathbf{e}_{{\sf{L}},k_{l}}$ follows $\mathbf{e}_{{\sf{L}},k_{l}}\sim \mathcal{CN}(\mathbf{0},\sigma_{e}^2\mathbf{I}_{N_{\sf{L}}})$. 

\begin{figure}
    \centering
    \includegraphics[width=0.85\linewidth]{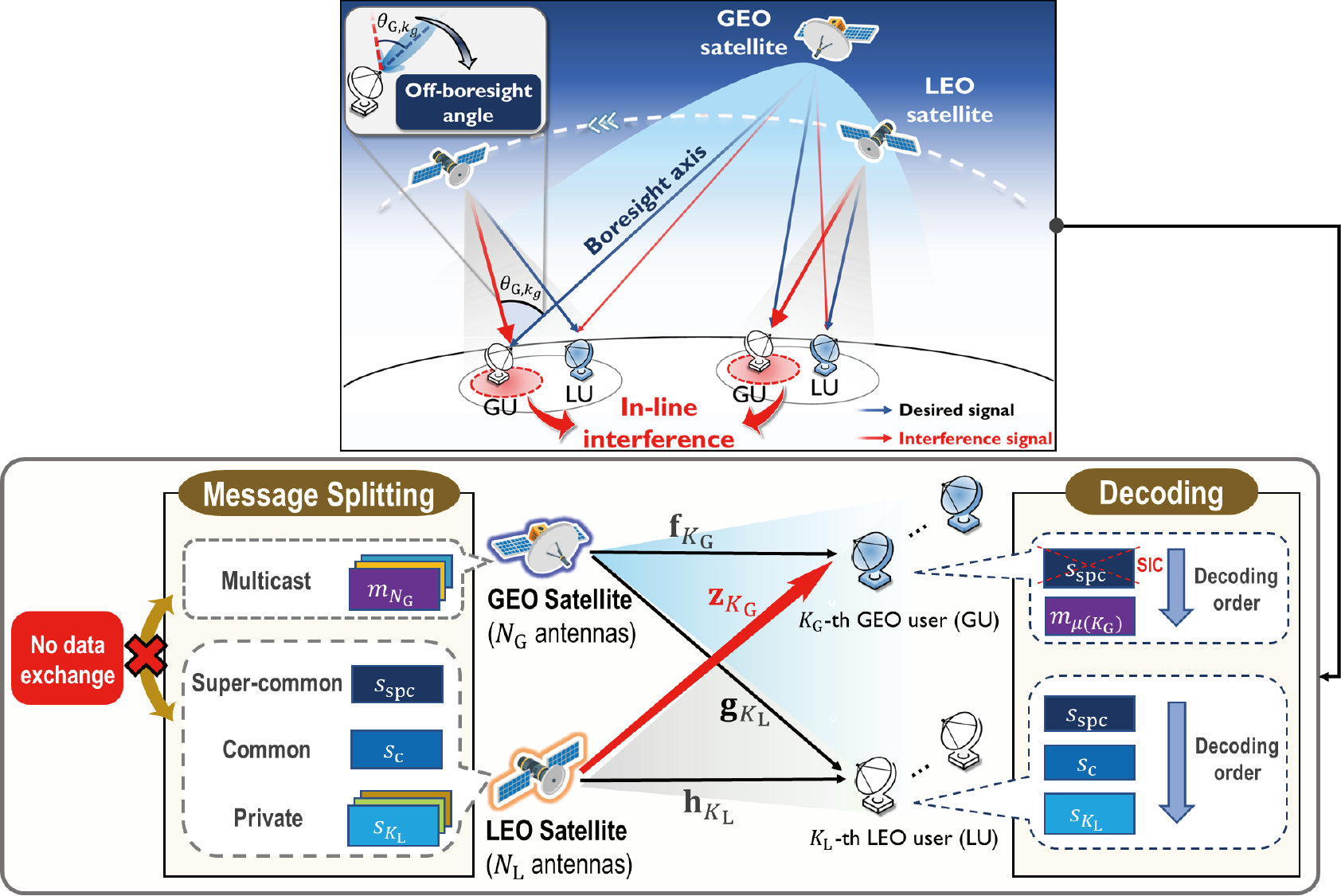}
    \caption{{GEO-LEO CSS system model based on RSMA.}}
    \label{fig:system model}
\end{figure}

%
As shown in Fig. \ref{fig:system model}, the boresight axis (i.e., the direction of maximum antenna gain for GU) refers to the direction from the GU to the GEO satellite, while the off-boresight angle, $\theta_{{\sf{G}},k_{g}}$, denotes the relative position of the LEO satellite from this axis \cite{Yunfeng:Access20}. 
%
As the off-boresight angle decreases (i.e., the LEO satellite moves close to the boresight axis), in-line interference can occur, resulting in the rapid increase of the IL from the LEO satellite to GU.
Therefore, the interference channel between the LEO satellite and $k_{g}$-th GU, $\mathbf{z}_{k_{g}} \in \mathbb{C}^{N_{\sf{L}} \times 1}$, depends on the receive antenna gain ${G_{\sf{R}}}(\theta_{{\sf{G}},k_{g}})$ \cite{ITU-R:S.1428-1}. 
{Under the imperfect CSI condition, the actual channel between the LEO satellite and the $k_g$-th GU can be formulated as $\mathbf{z}_{k_g} = \hat{\mathbf{z}}_{k_{g}} + \mathbf{e}_{{\sf{L}},k_g}$, where $\hat{\mathbf{z}}_{k_{g}}$ is the estimated channel and the channel estimation error $\mathbf{e}_{{\sf{L}},k_{g}}$ follows $\mathbf{e}_{{\sf{L}},k_{g}}\sim \mathcal{CN}(\mathbf{0},\sigma_{e}^2\mathbf{I}_{N_{\sf{L}}})$.}

{The channel vector between the GEO satellite and $k_{g}$-th GU can be expressed as 
{
\begin{equation}
\mathbf{f}_{k_{g}}=\sqrt{\left(\frac{\lambda}{4 \pi d_{{\sf{G}},k_{g}}}\right)^2 \frac{G_{\sf{R}}^{\sf{max}}}{\kappa B T}} \mathbf{b}_{{\sf{G}},k_{g}}^{\frac{1}{2}} \odot e^{j {\phi}_{{\sf{G}},k_{g}}},
\end{equation}}where $\left(\lambda / 4 \pi d_{{\sf{G}},k_{g}}\right)^2$ is the effect of free space path loss between GEO satellite and $k_{g}$-th GU. 
$\mathbf{b}_{{\sf{G}},k_{g}} =[b_{k_g,1},\ldots,b_{k_g,N_{\sf{G}}}]^T \in \mathbb{C}^{N_{\sf{G}} \times 1}$ is radiation pattern for $N_{\sf{G}}$ antenna feeds. 
$\mathbf{g}_{k_l} \in \mathbb{C}^{N_{\sf{G}} \times 1}$ is the channel vector between the GEO satellite and the ${k}_{l}$-th LU.
Similarly, we account for imperfect CSI in the GEO satellite channel model. Notably, the ultra-high mobility of LEO satellites (approximately 7.5 km/s) brings severe challenges on highly imperfect CSI at transmitters compared to GEO satellites which are stationary relative to the Earth.}

\subsection{Signal Model and Achievable Rate}
{We propose a novel RSMA scheme with a super-common message for GEO-LEO CSS.} The main goal of introducing the super-common message is to control the interference from the LEO satellite to the GU in compliance with ITU regulations \cite{ITU_R:Article}. The received signals at the ${k}_{l}$-th LU and ${k}_{g}$-th GU can be expressed as

{\begin{equation}
y_{k_{l}}^{\sf{L}}= \mathbf{h}_{k_{l}}^{H} \mathbf{x}_{\sf{L}} + \mathbf{g}_{k_{l}}^{H} \mathbf{x}_{\sf{G}} + n_{k_{l}}, \quad y_{k_{g}}^{\sf{G}}= \mathbf{f}_{k_{g}}^{H} \mathbf{x}_{\sf{G}} + \mathbf{z}_{k_{g}}^{H} \mathbf{x}_{\sf{L}} + n_{k_{g}},
 \label{eq:1}
\end{equation}}where $\mathbf{x}_{\sf{L}}$ and $\mathbf{x}_{\sf{G}}$ are the transmitted signals at the LEO satellite and GEO satellite, respectively. 
$n_{k_{l}} \sim \mathcal{CN}(0,\sigma_{n,k_{l}}^{2})$ and $n_{k_{g}} \sim \mathcal{CN}(0,\sigma_{n,k_{g}}^{2})$ are additive white Gaussian noise (AWGN). 
Since the satellite channel is normalized by the noise power $\kappa B T$, each user's noise variance is equal, $\sigma_{n,k_{l}}^{2} = \sigma_{n,k_{g}}^{2} = \sigma_{n}^{2} = 1$. 
{Note that the multicast message $W_{{\sf{m}},n_g}, \forall n_g \in \{1,\ldots,N_{\sf{G}}\}$, for $n_g$-th GEO beam is encoded into $m_{n_g} \sim \mathcal{CN}(0,1)$, which is transmitted by the GEO satellite.}
The unicast message $W_{k_l}$ for the ${k}_{l}$-th LU is divided into the super-common part, the common part, and the private part, that is, $W_{k_l}\rightarrow\{W_{{\sf{spc}},k_l}, W_{{\sf{c}},k_l}, W_{{\sf{p}},k_l}\}$, $\forall k_l \in \mathcal{K_L}$. 
All super-common parts are merged as $W_{\sf{spc}}$ and encoded into ${s}_{\sf{spc}}$ which is designed to be decodable by all GUs and LUs. 
Similarly, all common parts are merged as $W_{\sf{c}}$ and the corresponding encoded ${s}_{\sf{c}}$ should be decodable at all LUs.
On the contrary, the $k_l$-th private part is encoded into individual private stream ${s}_{k_l}$, $\forall k_l \in \mathcal{K_L}$, intended to be decodable at the $k_l$-th LU. 
We assume that the vector of symbol streams from the LEO satellite, $\mathbf{s} = \left[{s}_{1},\ldots,{s}_{K_{\sf{L}}},{s}_{\sf{c}},{s}_{\sf{spc}}\right]^T \in \mathbb{C}^{(K_{\sf{L}}+2) \times 1}$, follows $\mathbb{E}\{\mathbf{s}\mathbf{s}^H\} = \mathbf{I}$. {Then, the transmitted signals from the LEO and GEO satellites can be represented as \cite{DVB-S2X} 
{
\begin{equation}
\mathbf{x}_{\sf{L}}=\mathbf{p}_{\sf{spc}} {s}_{\sf{spc}} + \mathbf{p}_{\sf{c}} {s}_{\sf{c}} + \sum_{k_l=1}^{K_{\sf{L}}} \mathbf{p}_{k_l} {s}_{k_l}, \quad \mathbf{x}_{\sf{G}}=\sum_{n_g=1}^{N_{\sf{G}}}\mathbf{w}_{n_g} {m}_{n_g},
\end{equation}}where {$\mathbf{p}_{\sf{spc}} \in \mathbb{C}^{N_{\sf{L}} \times 1}, \mathbf{p}_{\sf{c}} \in \mathbb{C}^{N_{\sf{L}} \times 1}, \mathbf{p}_{k_l} \in \mathbb{C}^{N_{\sf{L}} \times 1}$,} and {$\mathbf{w}_{n_g} \in \mathbb{C}^{N_{\sf{G}} \times 1}$} are the precoding vectors for ${s}_{\sf{spc}}$, ${s}_{\sf{c}}$, ${s}_{k_l}$, and ${m}_{n_g}$, respectively. The transmitted power constraints at LEO and GEO satellites are expressed as {$\|\mathbf{p}_{\sf{spc}}\|^{2} + \|\mathbf{p}_{\sf{c}}\|^{2} + \sum_{k_l \in \mathcal{K_L}}\|\mathbf{p}_{k_l}\|^{2} \le P_{\sf{L}}$} and {$\sum_{n_g \in \mathcal{N_G}}\|\mathbf{w}_{n_g}\|^{2} \le P_{\sf{G}}$}, respectively.}

{Taking into account the imperfect CSI, the received signal}
{in \eqref{eq:1} can be reformulated as}\footnote{The coverage area of a GEO beam (radius: $140\sim225$ km) is substantially larger than that of a LEO beam (radius: $25\sim45$ km). Moreover, due to the high altitude of the GEO satellite, the power of the GEO channel is 20$\sim$30 dB lower than that of the LEO channel, which results in a weak interference leakage level from the GEO satellite to LEO users \cite{Wang:ICCC18}. Thus, it is sufficient to focus on interference from the single GEO beam.}
{
\begin{align}
    y_{k_{l}}^{\sf{L}} &= \hat{\mathbf{h}}_{k_{l}}^{H} \mathbf{p}_{\sf{spc}} s_{\sf{spc}} + \mathbf{e}_{{\sf{L}},k_{l}}^{H}\mathbf{p}_{\sf{spc}} s_{\sf{spc}} + \hat{\mathbf{h}}_{k_{l}}^{H} \mathbf{p}_{\sf{c}} s_{\sf{c}} + \mathbf{e}_{{\sf{L}},k_{l}}^{H}\mathbf{p}_{\sf{c}} s_{\sf{c}} 
    \nonumber \\ &+ \sum_{j=1}^{K_{\sf{L}}} \left( \hat{\mathbf{h}}_{k_{l}}^{H}\mathbf{p}_{j} s_{j} + \mathbf{e}_{{\sf{L}},k_{l}}^{H}\mathbf{p}_{j} s_{j}\right) + {\mathbf{g}_{k_{l}}^{H} \mathbf{w}_{\mu(k_l)} {m}_{\mu(k_l)}} + n_{k_{l}},
 \label{eq:2}
\end{align}
\begin{align}
    y_{k_{g}}^{\sf{G}} &= {\mathbf{f}_{k_{g}}^{H} \mathbf{w}_{\mu(k_g)} {m}_{\mu(k_g)}} + \hat{\mathbf{z}}_{k_{g}}^{H} \mathbf{p}_{\sf{spc}} s_{\sf{spc}} + \mathbf{e}_{{\sf{G}},k_{g}}^{H}\mathbf{p}_{\sf{spc}} s_{\sf{spc}} + \hat{\mathbf{z}}_{k_{g}}^{H} \mathbf{p}_{\sf{c}} s_{\sf{c}}  
    \nonumber \\ &+ \mathbf{e}_{{\sf{G}},k_{g}}^{H}\mathbf{p}_{\sf{c}} s_{\sf{c}} + \sum_{j=1}^{K_{\sf{L}}} \left( \hat{\mathbf{z}}_{k_{g}}^{H}\mathbf{p}_{j} s_{j} + \mathbf{e}_{{\sf{G}},k_{g}}^{H}\mathbf{p}_{j} s_{j}\right) + n_{k_{g}},
 \label{eq:3}
\end{align}}{where $\mu(k_l)$ denotes the mapping function that associates the $k_l$-th LU with the GEO beam where it is located \cite{Huanxi:WCL23}.}
With the imperfect CSIT and CSIR, the generalized mutual information (GMI) can be applied to determine the achievable rate \cite{Byungju:TVT23}. From \eqref{eq:2}, \eqref{eq:3}, the signal-to-interference-plus-noise ratio (SINR) for the super-common stream at the $k_l$-th LU and $k_g$-th GU can be expressed as
{
\begin{equation}
{\gamma}_{{\sf{spc}},k_{l}}^{\sf{L}} = \frac{|\hat{\mathbf{h}}_{k_l}^{H} \mathbf{p}_{{\sf{spc}}}|^{2}}{u_{k_l} + |\hat{\mathbf{h}}_{k_l}^{H} \mathbf{p}_{\sf{c}}|^{2} + \sum_{j=1}^{K_{\sf{L}}} |\hat{\mathbf{h}}_{k_l}^{H}\mathbf{p}_{j}|^{2} + \sigma_{n}^{2}},
\end{equation}
\begin{equation}
{\gamma}_{{\sf{spc}},k_{g}}^{\sf{G}} = \frac{|\hat{\mathbf{z}}_{k_g}^{H} \mathbf{p}_{\sf{spc}}|^{2}}{w_{k_g} + |\hat{\mathbf{z}}_{k_g}^{H} \mathbf{p}_{\sf{c}}|^{2} + \sum_{j=1}^{K_{\sf{L}}} |\hat{\mathbf{z}}_{k_g}^{H}\mathbf{p}_{j}|^{2} + \sigma_{n}^{2}}, \label{eq:4}
\end{equation}
}where {$u_{k_l} \triangleq {|\mathbf{g}_{k_l}^{H} \mathbf{w}_{\mu(k_l)}|^{2}} + \mathbb{E}[|\mathbf{e}_{{\sf{L}},k_l}^{H} \mathbf{p}_{\sf{spc}}|^{2}] + \mathbb{E}[|\mathbf{e}_{{\sf{L}},k_l}^{H} \mathbf{p}_{\sf{c}}|^{2}] + \sum_{j=1}^{K_{\sf{L}}} \mathbb{E}[|\mathbf{e}_{{\sf{L}},k_l}^{H} \mathbf{p}_{j}|^{2}]$} and {$w_{k_g} \triangleq {|\mathbf{f}_{k_g}^{H} \mathbf{w}_{\mu(k_g)}|^{2}} + \mathbb{E}[|\mathbf{e}_{{\sf{G}},k_g}^{H} \mathbf{p}_{\sf{spc}}|^{2}] + \mathbb{E}[|\mathbf{e}_{{\sf{G}},k_g}^{H} \mathbf{p}_{\sf{c}}|^{2}] + \sum_{j=1}^{K_{\sf{L}}} \mathbb{E}[|\mathbf{e}_{{\sf{G}},k_g}^{H} \mathbf{p}_{j}|^{2}]$}.

To protect the existing GEO system, we introduce the constraint of IL from LEO satellite to GUs in \eqref{eq:4} as follows 
{
\begin{equation}
     |\mathbf{z}_{k_g}^{H} \mathbf{p}_{\sf{c}}|^{2} + \sum_{j=1}^{K_{\sf{L}}} |\mathbf{z}_{k_g}^{H}\mathbf{p}_{j}|^{2} \le I_{\sf th}, \quad \forall {k_g}\in \mathcal{K_G},
    \label{eq:5}
\end{equation}}where $I_{\sf th}$ is the interference threshold level for GU. 
It is worth noting that due to the constraint \eqref{eq:5}, a large amount of power is allocated to the super-common message to enable the GU to mitigate the in-line interference.
Then, the achievable rates for $s_{\sf{spc}}$ at $k_l$-th LU and $k_g$-th GU are expressed by {$R_{{\sf{spc}},k_{l}}^{\sf{L}} = \log_{2} (1 + {\gamma}_{{\sf{spc}},k_{l}}^{\sf{L}})$} and {$R_{{\sf{spc}},k_{g}}^{\sf{G}} = \log_{2} (1 + {\gamma}_{{\sf{spc}},k_{g}}^{\sf{G}})$}, respectively.
{Finally, since all GUs and LUs need to decode $s_{\sf{spc}}$, the achievable rate for $s_{\sf{spc}}$ is defined as}

{ 
\begin{equation}
    R_{\sf{spc}} \triangleq \min \left(\min_{k_l \in \mathcal{K_L}} R_{{\sf{spc}},k_{l}}^{\sf{L}}, \min_{k_g \in \mathcal{K_G}} R_{{\sf{spc}},k_{g}}^{\sf{G}} \right) = \sum_{k_l=1}^{K_{\sf{L}}} C_{{\sf{spc}},k_l},
\end{equation}
}{where $C_{{\sf{spc}},k_l}$ denotes the portion of super-common rate for $W_{{\sf{spc}},k_l}$}.
After the SIC, the SINRs for $s_{\sf{c}}$ and $s_{k_l}$ at the $k_l$-th LU can be expressed as
{
\begin{equation}
    {\gamma}_{{\sf{c}},k_{l}} = \frac{|\hat{\mathbf{h}}_{k_l}^{H} \mathbf{p}_{\sf{c}}|^{2}} {u_{k_l} + \sum_{j=1}^{K_{\sf{L}}} |\hat{\mathbf{h}}_{k_l}^{H}\mathbf{p}_{j}|^{2} + \sigma_{n}^{2}},
\end{equation}
\begin{equation}
    \quad \quad {\gamma}_{{\sf{p}},k_{l}} = \frac{|\hat{\mathbf{h}}_{k_l}^{H} \mathbf{p}_{k_l}|^{2}} {u_{k_l} + \sum_{j=1,j \neq {k_l}}^{K_{\sf{L}}} |\hat{\mathbf{h}}_{k_l}^{H}\mathbf{p}_{j}|^{2} + \sigma_{n}^{2}}.
\end{equation}}
The achievable rates for $s_{\sf{c}}$ and $s_{k_l}$ at the $k_l$-th LU are expressed by {$R_{{\sf{c}},k_{l}} = \log_{2} (1 + {\gamma}_{{\sf{c}},k_{l}})$} and {$R_{{\sf{p}},k_{l}} = \log_{2} (1 + {\gamma}_{{\sf{p}},k_{l}})$}, respectively.
{As mentioned, since all LUs require decoding of the $s_{\sf{c}}$, the achievable rate for $s_{\sf{c}}$ is derived by}
{
\begin{equation}
    R_{\sf{c}} \triangleq \min_{k_l \in \mathcal{K_L}} R_{{\sf{c}},k_{l}} = \sum_{k_l=1}^{K_{\sf{L}}} C_{k_l},
\end{equation}
}{where $C_{k_l}$ denotes the portion of common rate for $W_{{\sf{c}},k_l}$}.

Thus, the total achievable rate for the $k_l$-th LU becomes
{
\begin{equation}
    R_{k_l} = C_{{\sf{spc}},k_l} + C_{k_l} + R_{{\sf{p}},k_l}.
\end{equation}
}

\section{The Proposed Traffic-Aware Throughput Maximization Framework} \label{SEC:Optimization}
{
Given heterogeneous traffic demands among users in practical satellite scenarios, the existing approaches of sum throughput maximization (STM) and MMF can lead to resource wastage.
This waste occurs when the throughput provided to certain users exceeds their traffic demand while simultaneously, there are users with unmet throughput.
Motivated by this, we aim to satisfy target traffic demands by minimizing the unmet throughput within a given satellite power budget.}
Therefore, the TTM problem can be formulated as follows:

{
\begin{align}
{\mathscr P_1:} \quad\quad& \max_{\mathbf{p}, \mathbf{c}_{\sf{spc}}, \mathbf{c}} \sum_{j=1}^{K_{\sf{L}}} \min \left(T_{j}, R_{j}\right) \\
\textrm{s.t.} & \quad R_{{\sf{spc}},k_{g}}^{\sf{G}} \ge \sum_{j=1}^{K_{\sf{L}}} C_{{\sf{spc}},j}, \quad \forall {k_g}\in \mathcal{K_G}, \\
& \quad |\mathbf{z}_{k_g}^{H} \mathbf{p}_{\sf{c}}|^{2} + \sum_{j=1}^{K_{\sf{L}}} |\mathbf{z}_{k_g}^{H}\mathbf{p}_{j}|^{2} \le I_{\sf th}, \quad \forall {k_g}\in \mathcal{K_G}, \\
& \quad R_{{\sf{spc}},k_{l}}^{\sf{L}} \ge \sum_{j=1}^{K_{\sf{L}}} C_{{\sf{spc}},j}, \quad R_{{\sf{c}},{k_l}} \ge \sum_{j=1}^{K_{\sf{L}}} C_{j}, \quad \forall {k_l}\in \mathcal{K_L}, \\
& \quad C_{{\sf{spc}},k_l} \ge 0, \quad C_{k_l} \ge 0, \quad \forall {k_l}\in \mathcal{K_L}, \quad \|\mathbf{p}\|^{2} \le P_{\sf{L}}, \label{eq:6}
%
\end{align}
}where $T_{k_l}$ represents the traffic demand for the $k_l$-th LU and {$\mathbf{p} =[\mathbf{p}_1^{H},\ldots,\mathbf{p}_{K_{\sf{L}}}^{H},\mathbf{p}_{\sf{c}}^{H},\mathbf{p}_{\sf{spc}}^{H}]^{H} \in \mathbb{C}^{N_{\sf{L}}(K_{\sf{L}}+2) \times 1}$. $\mathbf{c}_{\sf{spc}} = [C_{{\sf{spc}},1}, C_{{\sf{spc}},2}, \ldots, C_{{\sf{spc}},K_{\sf{L}}}]^T$} and {$\mathbf{c} = [C_1, C_2, \ldots, C_{K_{\sf{L}}}]^T$} are the rate portions of the super-common rate and common rate. 
%

%
%
{We first transform both the numerator and denominator of the achievable rate for each stream into quadratic forms as}
{
\begin{align}    
    &R_{{\sf{spc}},k_{g}}^{\sf{G}}=\log_{2}\left(\frac{\bar{\mathbf{p}}^{H}\mathbf{A}_{k_g}\bar{\mathbf{p}}}{{\bar{\mathbf{p}}}^{H}\mathbf{B}_{k_g}\bar{\mathbf{p}}}\right),
    R_{{\sf{spc}},k_{l}}^{\sf{L}}=\log_{2}\left(\frac{\bar{\mathbf{p}}^{H}\mathbf{D}_{k_l}\bar{\mathbf{p}}}{\bar{\mathbf{p}}^{H}\mathbf{F}_{k_l}\bar{\mathbf{p}}}\right), \label{eq:R_spc}\\
    &R_{{\sf{c}},k_{l}}=\log_{2}\left(\frac{\bar{\mathbf{p}}^{H}\mathbf{F}_{k_l}\bar{\mathbf{p}}}{\bar{\mathbf{p}}^{H}\mathbf{Q}_{k_l}\bar{\mathbf{p}}}\right),\quad R_{{\sf{p}},k_{l}}=\log_{2}\left(\frac{\bar{\mathbf{p}}^{H}\mathbf{Q}_{k_l}\bar{\mathbf{p}}}{\bar{\mathbf{p}}^{H}\mathbf{V}_{k_l}\bar{\mathbf{p}}}\right), \label{eq:R_c}
\end{align}}where {$\bar{\mathbf{p}}=[1,\mathbf{p}^H]^H$}, the matrices {$\mathbf{A}_{k_g}$, $\mathbf{B}_{k_g}$, $\mathbf{D}_{k_l}$, $\mathbf{F}_{k_l}$, $\mathbf{Q}_{k_l}$}, and {$\mathbf{V}_{k_l} \in \mathbb{C}^{(N_{\sf{L}}(K_{\sf{L}}+2)+1) \times (N_{\sf{L}}(K_{\sf{L}}+2)+1)}$} are positive definite and block-diagonal as follows:
{
\begin{align}    
    \mathbf{A}_{k_g}&=\textrm{blkdiag}({|\mathbf{f}_{k_g}^{H} \mathbf{w}_{\mu(k_g)}|^{2}},\hat{\mathbf{z}}_{k_g}\hat{\mathbf{z}}_{k_g}^{H},\hat{\mathbf{z}}_{k_g}\hat{\mathbf{z}}_{k_g}^{H},\ldots,\hat{\mathbf{z}}_{k_g}\hat{\mathbf{z}}_{k_g}^{H}) \nonumber 
    \\ &+ \textrm{blkdiag}(\sigma_{n}^{2},\sigma_{e}^{2}\mathbf{I}_{N_{\sf{L}}},\ldots,\sigma_{e}^{2}\mathbf{I}_{N_{\sf{L}}}),
    \\ \mathbf{B}_{k_g} &= \mathbf{A}_{k_g} - \textrm{blkdiag}(0,\mathbf{0},\ldots,\hat{\mathbf{z}}_{k_g}\hat{\mathbf{z}}_{k_g}^{H}),
    \\ \mathbf{D}_{k_l}&=\textrm{blkdiag}({|\mathbf{g}_{k_l}^{H} \mathbf{w}_{\mu(k_l)}|^{2}},\hat{\mathbf{h}}_{k_l}\hat{\mathbf{h}}_{k_l}^{H},\hat{\mathbf{h}}_{k_l}\hat{\mathbf{h}}_{k_l}^{H},\ldots,\hat{\mathbf{h}}_{k_l}\hat{\mathbf{h}}_{k_l}^{H}) \nonumber 
    \\ &+ \textrm{blkdiag}(\sigma_{n}^{2},\sigma_{e}^{2}\mathbf{I}_{N_{\sf{L}}},\ldots,\sigma_{e}^{2}\mathbf{I}_{N_{\sf{L}}}),
    \\ \mathbf{F}_{k_l} &= \mathbf{D}_{k_l} - \textrm{blkdiag}(0,\mathbf{0},\cdots,\hat{\mathbf{h}}_{k_l}\hat{\mathbf{h}}_{k_l}^{H}),
    \\ \mathbf{Q}_{k_l} &= \mathbf{F}_{k_l} - \textrm{blkdiag}(0,\mathbf{0},\ldots,\hat{\mathbf{h}}_{k_l}\hat{\mathbf{h}}_{k_l}^{H},\mathbf{0}),
    \\ \mathbf{V}_{k_l} &= \mathbf{Q}_{k_l} - \textrm{blkdiag}(0,\mathbf{0},\ldots,\underbrace{\hat{\mathbf{h}}_{k_l}\hat{\mathbf{h}}_{k_l}^{H}}_{(k_l+1)\textrm{-th block}},\ldots,\mathbf{0}).
\end{align}
}

{Note that since the rates are the difference of concave functions, $\mathscr P_1$ is a non-convex problem. 
Therefore, $\mathscr P_1$ requires reformulation into a convex problem.}
Further, we employ an alternating optimization algorithm by reformulating rates into a convex form and then deriving the upper and lower bounds of \eqref{eq:R_spc}, \eqref{eq:R_c} using exponential terms with auxiliary variables {$a_{k_g}, b_{k_g},  d_{k_l}, f_{1,k_l}, f_{2,k_l}, q_{1,k_l}, q_{2,k_l}, v_{k_l}$}:
{
\begin{flalign}
    &\bar{\mathbf{p}}^{H}\mathbf{A}_{k_g}\bar{\mathbf{p}}\! \ge \! e^{a_{k_g}}, \!\,
    \bar{\mathbf{p}}^{H}\mathbf{D}_{k_l}\bar{\mathbf{p}}\! \ge\! e^{d_{k_l}}, \!\,
    e^{q_{1,k_l}}\! \ge\! \bar{\mathbf{p}}^{H}\mathbf{Q}_{k_l}\bar{\mathbf{p}}\! \ge \!e^{q_{2,k_l}}\!,\label{eq:7}
    \\ 
    &\bar{\mathbf{p}}^{H}\mathbf{B}_{k_g}\bar{\mathbf{p}}\! \le\! e^{b_{k_g}},\!\,
    e^{f_{2,k_l}} \!\le\! \bar{\mathbf{p}}^{H}\mathbf{F}_{k_l}\bar{\mathbf{p}} \!\le \!e^{f_{1,k_l}}, \!\, \bar{\mathbf{p}}^{H}\mathbf{V}_{k_l}\bar{\mathbf{p}} \!\le \!e^{v_{k_l}}\!. \label{eq:8}
\end{flalign}
}From \eqref{eq:7}--\eqref{eq:8}, the lower bounds of rates can be derived as
{
\begin{align}
    &R_{{\sf{spc}},k_{g}}^{\sf{G}} \ge \frac{1}{\ln 2} (a_{k_g} - b_{k_g}), \quad   R_{{\sf{spc}},k_{l}}^{\sf{L}} \ge \frac{1}{\ln 2} (d_{k_l} - f_{1,k_l}), \\
    &R_{{\sf{c}},k_{l}} \ge \frac{1}{\ln 2} (f_{2,k_l} - q_{1,k_l}), \quad R_{{\sf{p}},k_{l}} \ge \frac{1}{\ln 2} (q_{2,k_l} - v_{k_l}).
\end{align}
}

To convert the non-convex quadratic constraints \eqref{eq:7}--\eqref{eq:8} into the convex forms, we apply the SDR technique \cite{Luo:SigMag10}.
Then, by replacing $\mathbf{X}=\mathbf{p}\mathbf{p}^H$ with the constraints $\mathbf{X} \succeq 0$ and $\textrm{rank}(\mathbf{X})=1$,  each quadratic term is transformed into
{
\begin{align}    
\bar{\mathbf{p}}^{H}\mathbf{A}_{k_g}\bar{\mathbf{p}} &
= {\sf tr}(\mathbf{A}_{k_g}\bar{\mathbf{p}}\bar{\mathbf{p}}^{H}) = {\sf tr}(\mathbf{A}_{k_g}\textrm{blkdiag}(1,\mathbf{X})).
    \label{eq:9}
\end{align} 
}

In particular, convexity can be achieved by relaxing the rank constraint in the SDR problem. 
{Since \eqref{eq:7} and \eqref{eq:8} consist of differences of convex functions, which are generally non-convex, we employ the CCCP} \cite{Thomas:Opt15}.
Using the first-order Taylor series approximation, the above constraints are derived as
{
\begin{align}
    &{\sf tr}(\mathbf{A}_{k_g}\textrm{blkdiag}(1,\mathbf{X})) \ge e^{a_{k_g}}, \,
    {\sf tr}(\mathbf{F}_{k_l}\textrm{blkdiag}(1,\mathbf{X})) \ge e^{f_{2,k_l}},
    \label{eq:10}
    \\ &{\sf tr}(\mathbf{D}_{k_l}\textrm{blkdiag}(1,\mathbf{X})) \ge e^{d_{k_l}}, \,
    {\sf tr}(\mathbf{Q}_{k_l}\textrm{blkdiag}(1,\mathbf{X})) \ge e^{q_{2,k_l}},
    \label{eq:11}
    \\ &{\sf tr}(\mathbf{B}_{k_g}\textrm{blkdiag}(1,\mathbf{X})) \le e^{b_{k_g}^{(m-1)}} (b_{k_g} - b_{k_g}^{(m-1)} + 1),
    \label{eq:12}
    \\ &{\sf tr}(\mathbf{F}_{k_l}\textrm{blkdiag}(1,\mathbf{X})) \le e^{f_{1,k_l}^{(m-1)}} (f_{1,k_l} - f_{1,k_l}^{(m-1)} + 1),
    \label{eq:13}
    \\ &{\sf tr}(\mathbf{Q}_{k_l}\textrm{blkdiag}(1,\mathbf{X})) \le e^{q_{1,k_l}^{(m-1)}} (q_{1,k_l} - q_{1,k_l}^{(m-1)} + 1),
    \label{eq:14}
    \\ &{\sf tr}(\mathbf{V}_{k_l}\textrm{blkdiag}(1,\mathbf{X})) \le e^{v_{k_l}^{(m-1)}} (v_{k_l} - v_{k_l}^{(m-1)} + 1), &&
    \label{eq:15} 
\end{align}
}where $m$ is the iteration number. {Similarly, the interference leakage constraint \eqref{eq:5} for GUs can be transformed as
{
\begin{equation}
{\sf tr}(\bar{\mathbf{B}}_{k_g}\mathbf{X}) \le I_{\sf{th}}, \quad \forall {k_g}\in \mathcal{K_G}, \label{eq:16} 
\end{equation}}where {$\bar{\mathbf{B}}_{k_g}\!=\!\textrm{blkdiag}(\hat{\mathbf{z}}_{k_g}\hat{\mathbf{z}}_{k_g}^{H} + \sigma_{e}^{2} \mathbf{I}_{N_{\sf{L}}},\ldots,\hat{\mathbf{z}}_{k_g}\hat{\mathbf{z}}_{k_g}^{H} + \sigma_{e}^{2}\mathbf{I}_{N_{\sf{L}}},\sigma_{e}^{2}\mathbf{I}_{N_{\sf{L}}})$}}, 
The reformulated optimization problem $\mathscr P_2$ at the $m$-th iteration is expressed as
\begin{align}
{\mathscr P_2:}\quad\quad& \max_{\substack{\mathbf{X}, \mathbf{c}_{\sf{spc}}, \mathbf{c}, \mathbf{a}, \mathbf{b}, \mathbf{d}, \\ \mathbf{f}_1, \mathbf{f}_2, \mathbf{q}_1, \mathbf{q}_2, \mathbf{v}}} \sum_{j=1}^{K_{\sf{L}}} \min \left(T_{j}, C_{{\sf{spc}},j} + C_{j} + \frac{q_{2,k_l} - v_{k_l}}{\ln 2} \right) \\
{\textrm{s.t.}}& \quad a_{k_g} - b_{k_g} \ge \sum_{j=1}^{K_{\sf{L}}} C_{{\sf{spc}},j} \ln 2, \quad \forall {k_g}\in \mathcal{K_G}, \\
%
%
& \quad d_{k_l} - f_{1,k_l} \ge \sum_{j=1}^{K_{\sf{L}}} C_{{\sf{spc}},j} \ln 2 , \quad \forall {k_l}\in \mathcal{K_L}, \\
& \quad f_{2,k_l} - q_{1,k_l} \ge \sum_{j=1}^{K_{\sf{L}}} C_{j} \ln 2, \quad \forall {k_l}\in \mathcal{K_L}, \\
%
%
& \quad C_{{\sf{spc}},k_l} \ge 0, \quad C_{k_l} \ge 0, \quad \forall {k_l}\in \mathcal{K_L}, \\
& \quad \mathbf{X} \succeq 0, \quad {\sf tr}(\mathbf{X}) \le P_{\sf{L}}, \quad \eqref{eq:10}-\eqref{eq:16}, &&
\end{align}where $\mathbf{a} \!=\! [a_{1},\ldots,a_{K_{\sf{G}}}]^T, \mathbf{b} \!=\! [b_{1},\ldots,b_{K_{\sf{G}}}]^T,
\mathbf{d} \!=\! [d_{1},\ldots,d_{K_{\sf{L}}}]^T,
\mathbf{f}_1 \!=\! [f_{1,1},\ldots,f_{1,K_{\sf{L}}}]^T, 
\mathbf{f}_2 \!=\! [f_{2,1},\ldots,f_{2,K_{\sf{L}}}]^T, 
\mathbf{q}_1 \!=\! [q_{1,1},\ldots,q_{1,K_{\sf{L}}}]^T, 
\mathbf{q}_2 \!=\! [q_{2,1},\ldots,q_{2,K_{\sf{L}}}]^T,
\mathbf{v} \!=\! [v_{1},\ldots,v_{K_{\sf{L}}}]^T$ are auxiliary variables.
The problem $\mathscr P_2$ is solved at each iteration to update $b_{k_{g}}^{(m-1)}, f_{1,k_{l}}^{(m-1)}, q_{1,k_{l}}^{(m-1)}, v_{k_{l}}^{(m-1)}$. 
{It should be noted that $\mathbf{X}^{\star}$ may not always satisfy the rank-1 constraint, $\textrm{rank}(\mathbf{X}^{\star})=1$. In such cases, $\mathbf{X}^{\star}$ is an infeasible solution to the original problem \cite{Luo:SigMag10}. 
To tackle this issue, we employ randomization to generate candidate precoding vectors.
However, approximate solutions are not always feasible for the interference leakage constraints for GU and the total power for the LEO satellite. 
Therefore, we propose Algorithm \ref{alg:TTM} to extract the feasible solution for the reformulated optimization problem.
In Algorithm \ref{alg:TTM}, ${D}^{(m)}, \epsilon,$ $M_1,$ and $M_2$ denote the value of {$\sum_{j=1}^{K_{\sf{L}}} \min \left(T_{j}, C_{{\sf{spc}},j} + C_{j} + (q_{2,k_l} - v_{k_l})/{\ln 2}\right)$} in the $m$-th iteration, the tolerance value, the maximum number of iterations, and the number of random vectors, respectively.
We first reconstruct the candidate precoding vector from the eigenvalue decomposition of $\mathbf{X}^{\star}$ and random vectors $\mathbf{f}$.
For each vector, we calculate the maximum value of interference leakage and the total power budget for the LEO satellite.
By flexibly rescaling the candidate precoding vectors to satisfy these constraints, we can finally obtain a feasible solution.

The proposed Algorithm \ref{alg:TTM} ensures convergence using the CCCP. 
This iterative method refines the solution at each step by solving a convex subproblem, achieving a local maximum \cite{Ying:TSP17}.
It starts from a feasible $(m-1)$-th solution and progressively improves $m$-th solution through this iterative process.}
%
%
\setlength{\textfloatsep}{0.3cm}
\begin{algorithm} [t]
\caption{SPC-RSMA TTM} \label{alg:TTM} 
\begin{algorithmic}
\State {Initialize}: $b_{k_{g}}^{(0)}, f_{1,k_{l}}^{(0)}, q_{1,k_{l}}^{(0)}, v_{k_{l}}^{(0)}, {{D}^{(0)}}, m \leftarrow 0.$
\State {\textbf{repeat}}
\State \hspace{0.1cm} $m \leftarrow m+1$ 
\State \hspace{0.1cm} Given $b_{k_{g}}^{(m-1)}, f_{1,k_{l}}^{(m-1)}, q_{1,k_{l}}^{(m-1)}, v_{k_{l}}^{(m-1)}$, solve ${\mathscr P_2}$
\State {\textbf{until}} $|{D}^{(m)}-{D}^{(m-1)}| \le \epsilon$ or $m \ge M_1$
\State Generate {$M_2$} random vectors $\mathbf{f} \sim \mathcal{CN}(0,\mathbf{I}_{N_{\sf{L}}(K_{\sf{L}}+2)})$
\State Decomposition of the optimal $\mathbf{X}^{\star}=\mathbf{U} \mathbf{\Sigma} \mathbf{U}^{H}$ from ${\mathscr P_2}$
\State {Obtain $\mathbf{p}=\mathbf{U}\mathbf{\Sigma}^{1/2}\mathbf{f}$ and the maximum IL for GU ($I_{\sf{max}}$)
\State $I_{\sf{max}} \leftarrow \text{max}_{k_g \in \mathcal{K_G}} |\mathbf{z}_{k_g}^{H} \mathbf{p}_{\sf{c}}|^{2} + \sum_{j=1}^{K_{\sf{L}}} |\mathbf{z}_{k_g}^{H}\mathbf{p}_{j}|^{2}$}
\If{$I_{\sf th} < I_{\sf{max}}$ or $\|\mathbf{p}\|^{2} > P_{\sf{L}}$}
    \State $\mathbf{p} \leftarrow \text{min}(\sqrt{I_{\sf th}/I_{\sf max}}, \sqrt{P_{\sf{L}}}/{\|\mathbf{p}\|})\mathbf{p}$
\EndIf
\State Choose the best $\mathbf{p}^{\star}$ to maximize LEO system throughput
\State {\textbf{output}}: $\mathbf{p}^{\star}, \bar{\mathbf{c}}_{\mathrm{spc}}^{\star}, \bar{\mathbf{c}}^{\star}$
\end{algorithmic}
\end{algorithm}

{\textbf{Remark 1. (Key difference from existing data sharing approach between GEO and LEO systems)}: \emph{Note that data sharing approaches \cite{Pengwenlong:TWC22,Yunnuo:Arxiv23} can provide seamless throughput for GEO-LEO CSS at the cost of coordination in the power control between GEO and LEO systems (i.e., considerable signaling overhead and tight synchronization). } 
\emph{
%
On the other hand, our framework enables interference management led by LEO satellites without the data exchange between GEO and LEO systems by sharing the CSI for all links via gateways corresponding to each system.}
} 

{\textbf{Remark 2. (Computational complexity of Algorithm \ref{alg:TTM})}: \emph{The computational complexity of the SDR problem depends on the dimension of matrix $\mathbf{X}$ and the number of linear matrix inequalities (LMIs).
%
In our approach, the dimension of $\mathbf{X}$, derived from the joint precoding vectors, mainly affects the computational complexity.
The worst-case computational complexity for each iteration step of Algorithm \ref{alg:TTM} is expressed as $\mathcal{O}([{N_{\sf{L}} K_{\sf{L}}}]^{3.5}\log(\varepsilon^{-1}))$, where $\varepsilon$ denotes the convergence tolerance} \cite{Luo:SigMag10}.}

%
\begin{table}[t]
\begin{center}
\caption{\begin{small}Simulation Parameters\end{small}} 
\label{Table1:Parameters}
\renewcommand{\arraystretch}{1.3} 
\begin{tabular}{ccc} \hline\hline
\multicolumn{1}{c|}{\textbf{Parameters}}                        & \multicolumn{1}{c|}{\textbf{LEO}} & \textbf{GEO}             
  \\ \hline
\multicolumn{1}{c|}{Carrier frequency}                          & \multicolumn{2}{c}{20 GHz}                                                          \\ \hline
\multicolumn{1}{c|}{Bandwidth}                          & \multicolumn{2}{c}{500 MHz}                                                          \\ \hline
\multicolumn{1}{c|}{Altitude of satellite} & \multicolumn{1}{c|}{600 km}     & 35,786 km                                                  \\ \hline
\multicolumn{1}{c|}{Satellite Tx maximum antenna gain} & \multicolumn{1}{c|}{30.5 dBi}     & 50.5 dBi                                                  \\ \hline
\multicolumn{1}{c|}{User Rx maximum antenna gain}      & \multicolumn{2}{c}{39.7 dBi}                                                                  \\ \hline
\multicolumn{1}{c|}{$I_{\sf{th}}, \, [T_{1}, T_{2}, T_{3}, T_{4}]$}            & \multicolumn{2}{c}{1, [1.5, 1, 2.5, 1.5]} \\ \hline
\multicolumn{1}{c|}{Maximum number of iteration}            & \multicolumn{2}{c}{$M_1=20$} \\ \hline
\multicolumn{1}{c|}{{Number of random vectors}}            & \multicolumn{2}{c}{$M_2=5\times10^{3}$} \\ \hline
\multicolumn{1}{c|}{Tolerance value}            & \multicolumn{2}{c}{$\epsilon=10^{-6}$} \\ \hline\hline
\end{tabular}
\end{center}
\end{table}
\section{Simulation results}\label{SEC:simulation}
In this section, we evaluate the numerical results to validate the proposed algorithm.
We deploy a GEO-LEO CSS where the GEO satellite equipped with $N_{\sf{G}}=2$ transmit antennas serves $K_{\sf{G}}=2$ GUs while the LEO satellite equipped with $N_{\sf{L}}=2$ transmit antennas serves $K_{\sf{L}}=4$ LUs. 
The simulation parameters are summarized in Table \ref{Table1:Parameters}. 
%
We assume that the LEO satellite operates two beams for all simulation cases. 
GUs are located at the midpoint between LEO beam centers, while LUs are identically spaced from the center of each beam.
%
In this case, LU 2 and LU 3 are located near the GUs.

We compare the proposed TTM based on RSMA with the super-common (SPC) message (denoted as {\sf{SPC-RSMA TTM}}) with the conventional techniques:
1) {\sf{RSMA TTM}} allocates zero power to the SPC part.
2) Spatial division multiple access {\sf{(SDMA) TTM}} allocates zero power to both the SPC and common parts.
3) {\sf{Progressive pitch TTM}} selectively deactivates a subset of LEO beams \cite{Ting:Int-J-SatCom-Net21}.
4) {\sf{Band splitting TTM}} orthogonally divides a frequency band into smaller sub-bands for each system \cite{Christophe:DySPAN19}.
Also, the proposed scheme is compared to two existing schemes:
5) {\sf{SPC-RSMA STM}} maximizes the sum throughput across all users \cite{Khan:TWC23}.
6) {\sf{SPC-RSMA MMF}} maximizes the minimum throughput among users \cite{Byungju:TVT23}.

%
\begin{figure}
    \centering
    \includegraphics[width=0.85\linewidth]{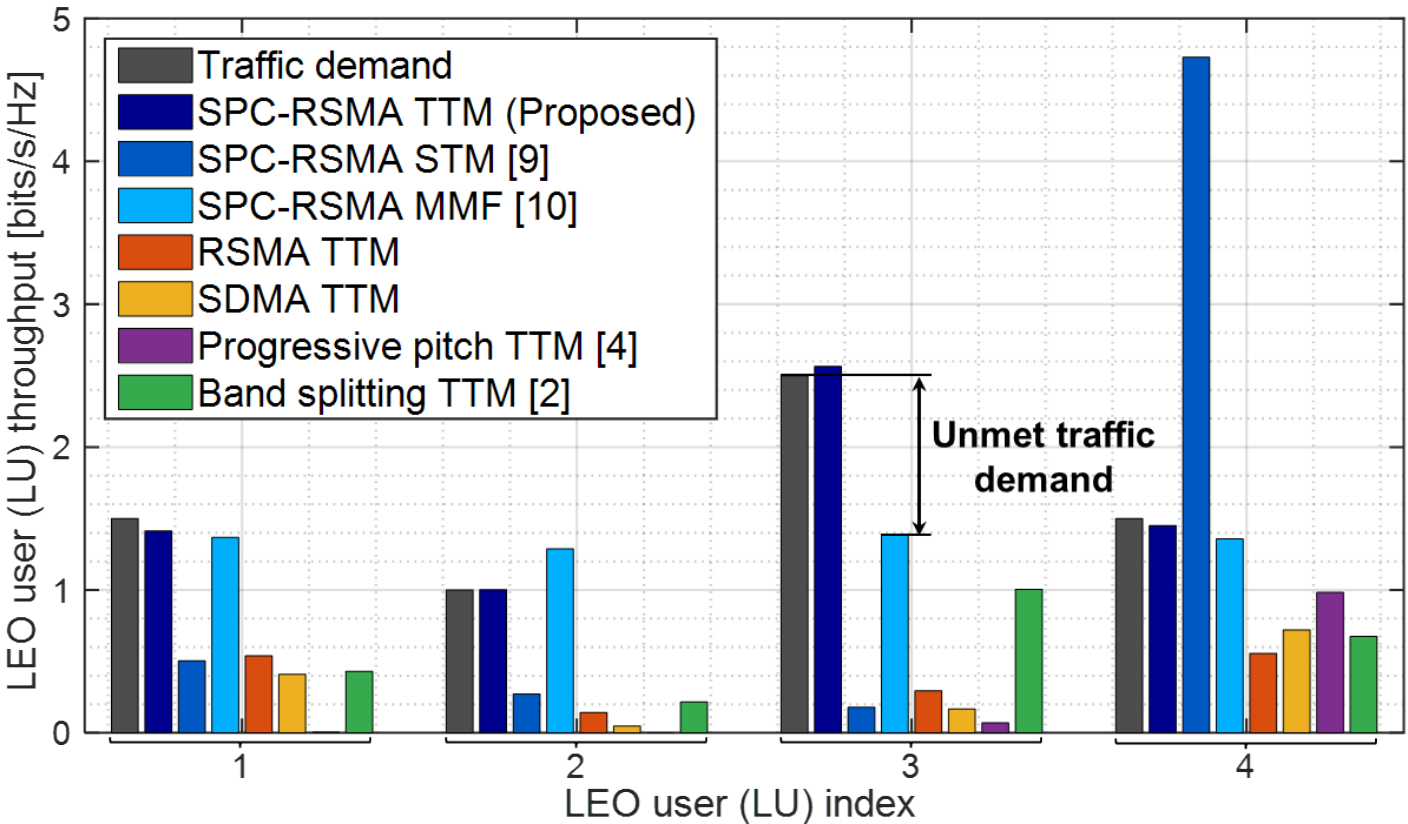} 
    \caption{LU throughput according to LU positions under imperfect CSIT and CSIR ($\sigma_{e}^2 = 0.05$).}
    \label{fig2:bar_graph}
\end{figure}
In Fig. \ref{fig2:bar_graph}, we compare the throughput of each LU for the proposed scheme with existing techniques when $\theta_{G,k_{g}} = 0.2^{\circ}$. 
%
We can observe that the proposed {\sf{SPC-RSMA TTM}} satisfies most of the traffic demand even with imperfect CSI. 
%
On the contrary, {\sf{SPC-RSMA STM}} concentrate the limited power to a few LUs with strong channels while {\sf{SPC-RSMA MMF}} improves fairness among all LUs; thus, both lead to unmet traffic demands of users.
%
Note that the conventional multiple access schemes show substantial LU throughput gaps from traffic demands. 
Since LU 2 and LU 3 are close to GUs, their throughputs are negligible to meet the IL constraint for GU.
Furthermore, since the activated beam for the progressive pitch scheme is assumed to target LU 3 and LU 4, the throughput is biased toward LU 4 located furthest from GUs.
In addition, the band splitting method provides service for LUs independent of GUs' locations, while it leads to an overall reduced throughput for each LU due to the splitting frequency resources.
We also observe that average achievements for the target traffic demands are 98\% for {\sf{SPC-RSMA TTM}}, 38\% for {\sf{SPC-RSMA STM}}, 88\% for {\sf{SPC-RSMA MMF}}, 24\% for {\sf{RSMA TTM}}, 20\% for {\sf{SDMA TTM}}, 17\% for {\sf{Progressive pitch TTM}}, and 34\% for {\sf{Band splitting TTM}}, respectively.

\begin{figure}[!ht]
    \centering
    \includegraphics[width=0.85\linewidth]{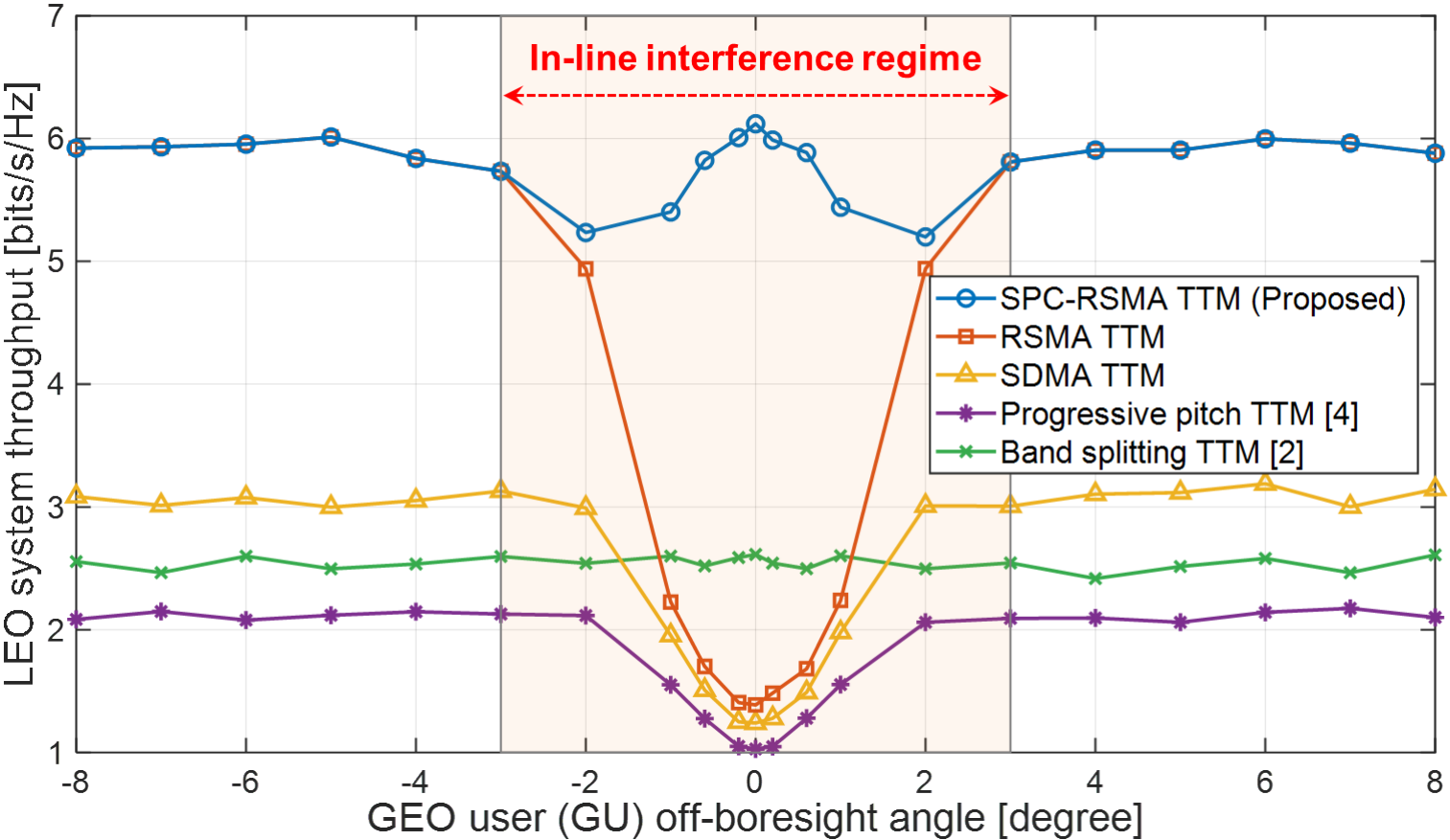} 
    \caption{LEO system throughput as a function of GU off-boresight angle under imperfect CSIT and CSIR ($\sigma_{e}^2 = 0.05$).}
    \label{fig3:off-boresight angle}
\end{figure}

Fig. \ref{fig3:off-boresight angle} shows the LEO system throughput versus GU off-boresight angle $\theta_{G,k_{g}}$.
For locations beyond the in-line interference regime (i.e., $|\theta_{G,k_{g}}| \geq 3^{\circ}$), the system throughput for each scheme is maintained at a certain level, respectively.
%
However, in the in-line interference regime (i.e., $|\theta_{G,k_{g}}| < 3^{\circ}$), to meet the IL constraint \eqref{eq:5}, the conventional techniques without the help of the SPC part cannot guarantee the LEO system throughput.
In contrast, the proposed {\sf{SPC-RSMA TTM}} flexibly adjusts the power allocated to the SPC part, ensuring LEO system throughput regardless of the relative position between the GUs and the LEO satellite.
In addition, {\sf{RSMA TTM}} achieves a better system throughput compared to {\sf{SDMA TTM}} and {\sf{Progressive pitch TTM}} techniques due to effective interference management among LUs.
Consequently, we observe that {\sf{SPC-RSMA TTM}} shows seamless connectivity even in the GEO-LEO in-line interference regime by achieving the limitation of the IL constraint for GU and the flexible interference management among the LUs.

\begin{figure}[!ht]
    \centering
    \includegraphics[width=0.85\linewidth]{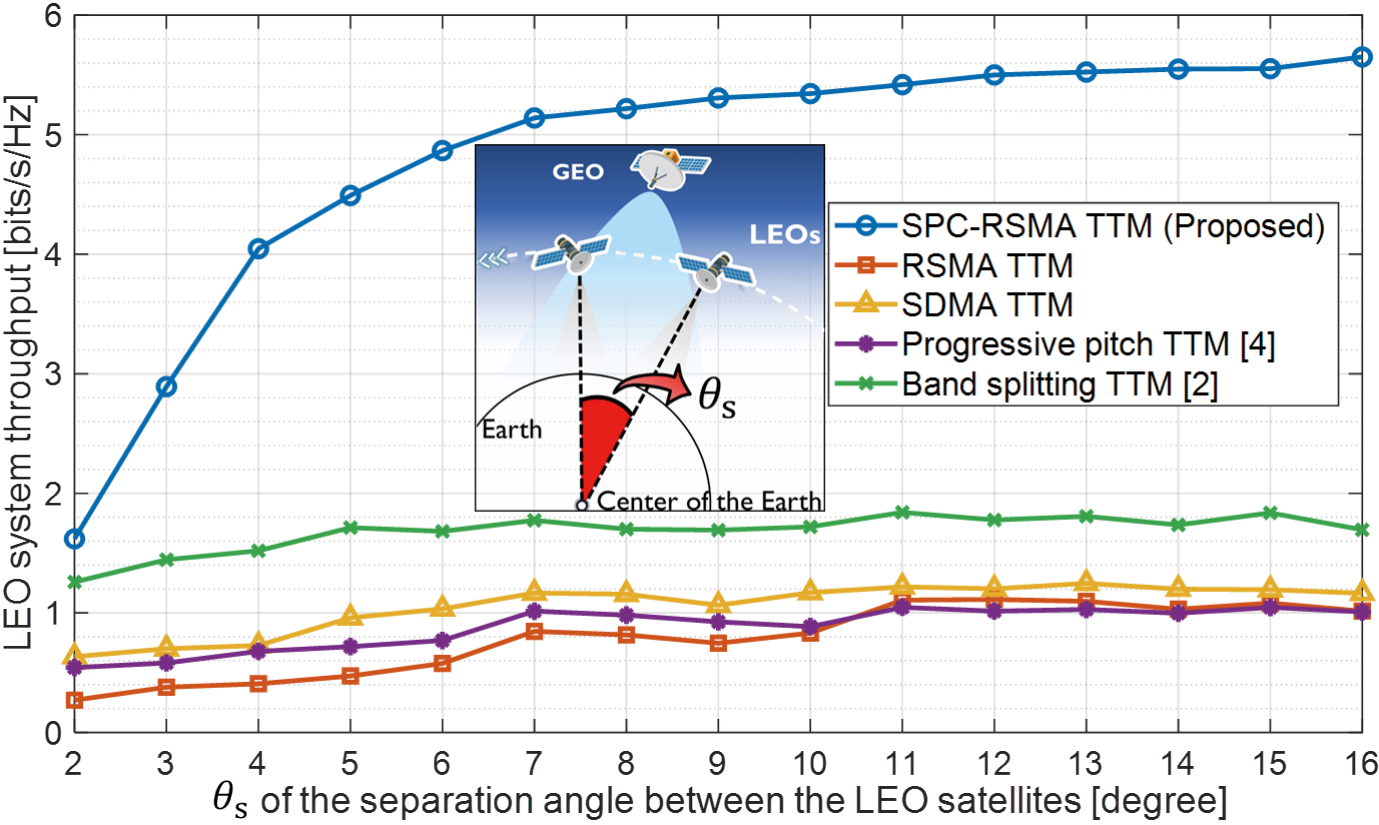} 
    \caption{LEO system throughput as a function of separation angle between the LEO satellites under imperfect CSIT and CSIR ($\sigma_{e}^2 = 0.05$).}
    \label{fig4:multiple LEO}
\end{figure}

To further validate the practical applicability of our proposed method, we extend the scenario that deploys a GEO-LEO CSS depicted in Fig. \ref{fig2:bar_graph}, by adding another LEO satellite into the same orbit.
Fig. \ref{fig4:multiple LEO} shows the LEO system throughput versus the separation angle between the LEO satellites with respect to the center of the Earth (i.e., $\theta_{\sf{s}}$).
While the additional LEO satellite leads to a decrease in LEO system throughput, the throughput loss is less than 10\% for $\theta_{\sf{s}} \!\geq\! 6^\circ$, even compared to a GEO-LEO CSS scenario.
Given that Starlink deploys 22 LEO satellites per orbit (i.e., $\theta_{\sf{s}}\!=\!16^\circ$) at altitudes of 540-550 km \cite{Starlink}, we verify that the proposed method can be applied to more realistic multi-layer satellite scenarios.
%

%
%

\section{Conclusion}\label{sec:conclude}
In this paper, we have investigated TTM based on RSMA with a super-common message for GEO-LEO CSS under imperfect CSIT and CSIR. 
To protect the GEO system, we have introduced a super-common message that allows GUs to mitigate in-line interference in an efficient manner.
Also, we have proposed the TTM to maximize the system throughput under various traffic demands of users.
The numerical results show that the proposed framework flexibly mitigates the in-line interference and guarantees seamless coverage for GEO-LEO CSS even in the in-line interference regime.

%

\ifCLASSOPTIONcaptionsoff
  \newpage
\fi

\end{document}